\documentclass[pra,aps,showpacs,groupedaddress,superscriptaddress,twocolumn,toc=flat]{revtex4-1}

\usepackage[colorlinks=true,linkcolor=blue,urlcolor=blue,citecolor=blue]{hyperref}

\usepackage[utf8x]{inputenc}
\usepackage{color}
\usepackage{bbm} 

\usepackage{amsfonts,amsmath,amssymb,stmaryrd}

\usepackage{graphicx}
\usepackage{subfigure}  
\usepackage{bbm} 
\usepackage{hyperref}
\usepackage{epsfig}
\usepackage{mathrsfs}
\usepackage{verbatim}
\usepackage{centernot}
\usepackage{ulem}
\usepackage{longtable}
\usepackage{nicefrac}
\usepackage[framemethod=tikz]{mdframed}
 
\renewcommand{\l}{\left(}
\renewcommand{\r}{\right)}

\newcommand{\bra}[1]{\langle#1|}
\newcommand{\ket}[1]{|#1\rangle}

\renewcommand{\ij}{{\langle \vec{i}, \vec{j} \rangle}}

\renewcommand{\H}{\hat{\mathcal{H}}}

\renewcommand{\c}{\hat{c}}

\newcommand{\cd}{\hat{c}^\dagger}
\newcommand{\rh}{\hat{\rho}}

\usepackage{array}

\usepackage{cancel,ifthen}
\newcommand{\cmnt}[2][NoInPuT]{\ifthenelse{\equal{#1}{NoInPuT}}{}{{\color{red}\sout{#1}}} {\color{blue} #2}}

\usepackage{bm}	
\renewcommand{\vec}[1]{\bm{#1}}

\LTcapwidth=\textwidth

\begin{document}
\normalem	

\title{Dynamical signatures of thermal spin-charge deconfinement in the doped Ising model}

\author{Lauritz Hahn}
\affiliation{Department of Physics and Arnold Sommerfeld Center for Theoretical Physics (ASC), Ludwig-Maximilians-Universit\"at M\"unchen, Theresienstr. 37, M\"unchen D-80333, Germany}
\affiliation{Munich Center for Quantum Science and Technology (MCQST), Schellingstr. 4, D-80799 M\"unchen, Germany}

\author{Annabelle Bohrdt}
\address{ITAMP, Harvard-Smithsonian Center for Astrophysics, Cambridge, MA 02138, USA}
\affiliation{Department of Physics, Harvard University, Cambridge, Massachusetts 02138, USA}


\author{Fabian Grusdt}
\email[Corresponding author email: ]{fabian.grusdt@physik.uni-muenchen.de}
\affiliation{Department of Physics and Arnold Sommerfeld Center for Theoretical Physics (ASC), Ludwig-Maximilians-Universit\"at M\"unchen, Theresienstr. 37, M\"unchen D-80333, Germany}
\affiliation{Munich Center for Quantum Science and Technology (MCQST), Schellingstr. 4, D-80799 M\"unchen, Germany}


\date{\today}

\begin{abstract}
The mechanism underlying charge transport in strongly correlated quantum systems, such as doped antiferromagnetic Mott insulators, remains poorly understood. Here we study the expansion dynamics of an initially localized hole inside a two-dimensional (2D) Ising antiferromagnet at variable temperature. Using a combination of classical Monte Carlo and a truncated basis method, we reveal two dynamically distinct regimes: A spin-charge confined region below a critical temperature $T^*$, characterized by slow spreading, and a spin-charge deconfined region above $T^*$, characterized by an unbounded diffusive expansion. The deconfinement temperature $T^*\approx 0.65 J_z$ we find is around the N\'eel temperature $T_{\rm N} = 0.567 J_z$ of the Ising background in 2D, but we expect $T^* < T_{\rm N}$ in higher dimensions. In both regimes we find that the mobile hole does not thermalize with the Ising spin background on the considered time scales, indicating weak effective coupling of spin- and charge degrees of freedom. Our results can be qualitatively understood by an effective parton model, and can be tested experimentally in state-of-the-art quantum gas microscopes.
\end{abstract}

\maketitle

\emph{Introduction.--}
In the field of high-$T_c$ superconductivity emerging from correlated insulating parent states \cite{Lee2006}, understanding the properties of individual charge carriers in doped 2D antiferromagnets (AFM) has been a central goal. While a magnetic, or spin-, polaron forms at low doping, experiments observe a cross-over from a polaronic metal at low doping to a Fermi liquid at high doping \cite{Koepsell2020_FL}. Although the ground state properties of magnetic polarons at low doping are essentially agreed upon \cite{Shraiman1988,Kane1989,Sachdev1989,Dagotto1990,Martinez1991,Liu1992,Brunner2000,Grusdt2019,nielsen2021spatial}, their fate at elevated temperatures or non-zero doping, as well as their far-from equilibrium dynamics, remains poorly understood.

Recently, ultracold atom experiments have ventured into this regime \cite{Bohrdt2021PWA}. In equilibrium, the dressing cloud of a magnetic polaron has been observed for the first time \cite{Koepsell2019}, and the dynamical spreading of an initially localized hole in 2D has revealed a significant slow-down associated with the presence of spin-correlations \cite{Ji2020}. Theo\-re\-ti\-cal work on the dynamical properties of doped holes has revealed signatures of parton \cite{Bohrdt2020_NJP} and string formation \cite{Mierzejewski2011,Golez2014,Bohrdt2020_NJP,Hubig2020,Bohrdt2020_ARPES} at low temperatures, and predicted diffusive or sub-diffusive spreading at infinite temperatures depending on the interactions between the spins \cite{Carlstrom2016PRL,kanasznagy2017,Bohrdt2020_NJP}.

\begin{figure}[t!]
\centering
  \includegraphics[width=0.99\linewidth]{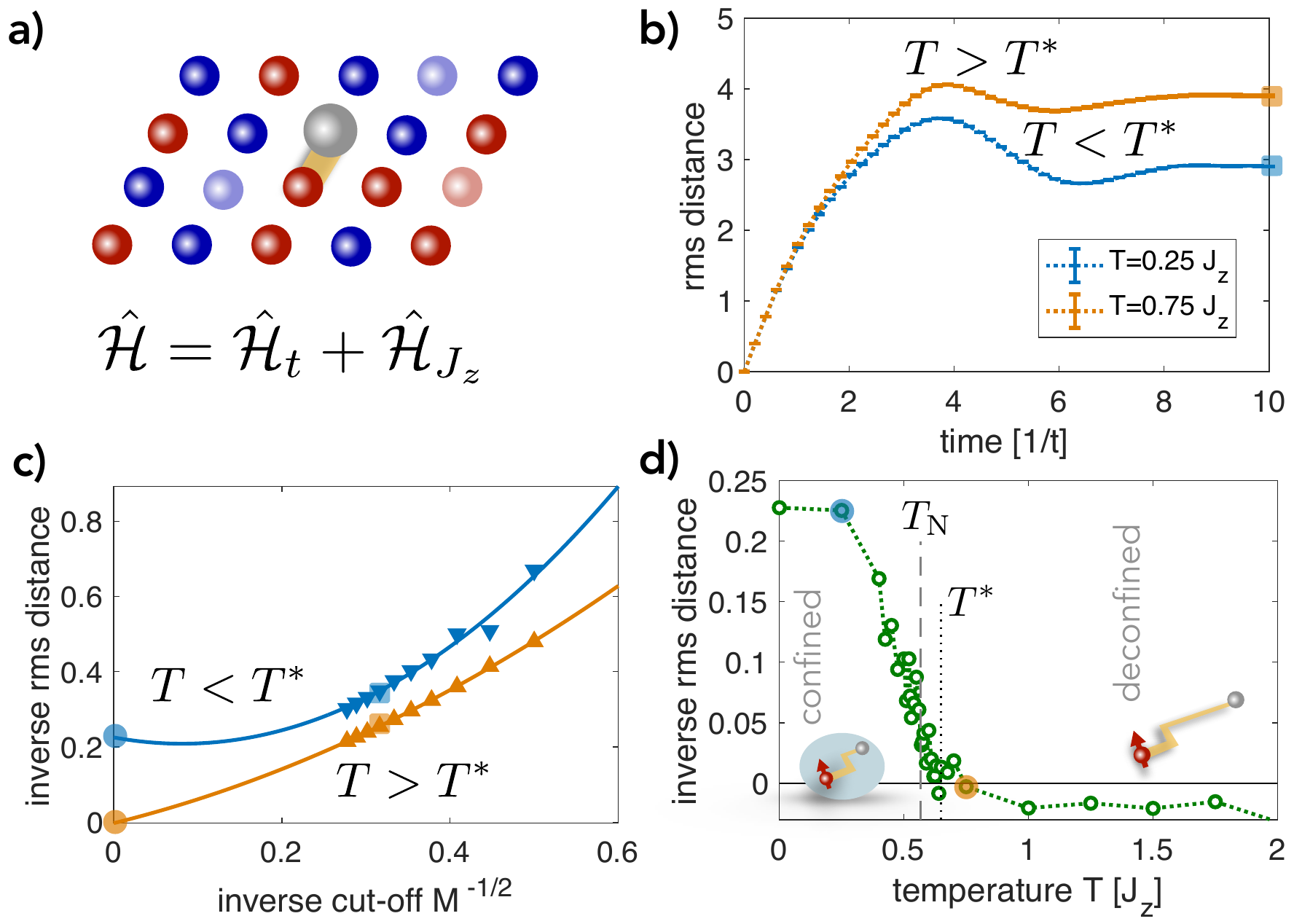}
\caption{We study the  spreading of an initially localized hole in a thermal Ising background at  temperatures $T$, a). The root mean square (rms) distance from the origin, shown for $M=10$, reveals slow (fast) spreading at low (high) temperatures, b). We study the long-time value of the inverse rms distance to extrapolate its value in the thermodynamic limit when our finite-size cut-off $M \to \infty$, c). Plotting the result for different temperatures, d), reveals a dynamically confined (deconfined) regime at low (high) temperatures. We show plots for $t/J_z = 3$; symbols in b)-d) correspond to same data.}
\label{fig1}
\end{figure}

Here we study the non-equilibrium dynamics of an initially localized single dopant in a thermal 2D Ising background, see Fig.~\ref{fig1}a). While previous studies addressed this problem in the limits of infinite temperature  with \cite{Bohrdt2020_NJP} or without \cite{Carlstrom2016PRL,kanasznagy2017} Ising interactions $J_z$, and at zero temperature with Ising couplings \cite{Grusdt2018tJz}, we systematically tune the temperature $T$ across the Ising critical point at $T_{\rm N} = 0.567 J_z$. Combining numerical Monte-Carlo and truncated basis methods, we reveal two regimes with qualitatively distinct hole dynamics, see Fig.~\ref{fig1}. By comparing our results to an effective parton model, we argue that the low-temperature behavior corresponds to spin-charge confinement, whereas spin and charge are deconfined at high-temperatures, see Fig.~\ref{fig1}d). 

Further, we study the thermalization dynamics of the mobile hole. We find that after a few tunneling times the hole quickly realizes a steady-state which differs significantly from the thermal state, especially in the deconfined regime at high temperatures. This finding is interesting since the $t-J_z$ model is neither believed to be integrable nor localizing \cite{Bulaevskii1968,Trugman1988}.

\emph{Model.--}
Some of the most relevant aspects of hole dynamics in AFM environments can be captured by the $t-J_z$ Hamiltonian \cite{Chernyshev1999,Grusdt2018tJz} in $d=2$ dimensions,
\begin{equation}
\H =-t\sum_{\ij,\sigma} \hat{\mathcal{P}} \l\cd_{\vec i,\sigma}\c_{\vec j,\sigma} +\text{h.c.}\r \hat{\mathcal{P}} +J_z\sum_{\ij}\hat{S}_{\vec{i}}^z \hat{S}_{\vec{j}}^z,
\end{equation}
where the first term $\H_{t}$ is the nearest-neighbor (NN) hopping with amplitude $t$, $\c_{\vec{j},\sigma}$ annihilates a fermion of spin $\sigma$ at site $\vec{j}$ and $\hat{\mathcal{P}}$ is a projector to the subspace without double occupancies. The second term $\H_{J_z}$ denotes NN AFM Ising interactions of the spins $\hat{S}^z_{\vec{j}} = \sum_\sigma (-1)^\sigma \cd_{\vec{j},\sigma} \c_{\vec{j},\sigma}$ with strength $J_z>0$. 

While the $t-J_z$ model constitutes a strong simplification, it captures several aspects relevant to experiments on strongly correlated electrons in cuprates, e.g. the formation of string patterns \cite{Bulaevskii1968,Grusdt2018tJz,Chiu2019Science}. Although the Ising background $\H_{J_z}$ itself is classical, the non-commuting hopping term $\H_t$ allows to couple most spin states to each other already by a single mobile dopant; even in a perfect N\'eel background at zero temperature, Trugman loops lead to coherent hole motion \cite{Trugman1988,Poilblanc1992,Grusdt2018tJz}. This renders $\H$ a truly quantum Hamiltonian. 

In the following, we study quantum quenches starting from an undoped thermal Ising state described by the density matrix $\rh_0 = e^{- \beta \H_{J_z}} / Z_0$, where $\beta = 1/T$ is the inverse of temperature $T$ and we set $k_B=1$. At time $\tau=0$ a single hole is created in the origin at $\vec{j}=\vec{0}$ and the initial state is $\rh(0) = \sum_\sigma \c_{\vec{0},\sigma} ~  \rh_0 ~ \cd_{\vec{0},\sigma}$.

\emph{Numerical technique.--}
To calculate the time-evolved density matrix $\rh(\tau)$ with a single hole, we leverage the classical nature of the Ising background $\H_{J_z}$. Specifically, we sample thermal initial spin states, dope them with one hole, and calculate their time-evolution by a truncated-basis method \cite{Mierzejewski2011,Vidmar2012,Golez2014}.
 
For a given eigenstate $\ket{\Psi^n}$ of the 2D Ising Hamiltonian, we obtain an initial one-hole state $\ket{\psi_1^n}=\c_{\vec{0},\sigma_{\vec{0}}} \ket{\Psi^n}$ by removing the fermion at the origin with spin $\sigma_{\vec{0}}$. Repeated applications of the terms in $\H_{t}$ then generate new states which we add to the truncated basis $\{\ket{\psi_\nu^n}\}_{\nu=1...d_M}$ used for numerical time evolution. In this process, orthonormality is guaranteed by projecting each new state onto all previous states. Since $\H_t$ is applied in each step, the total number of iterations $M$ corresponds to the largest number of hops the hole can perform in the truncated basis without retracing its path; the dimension $d_M$ of the truncated basis grows exponentially with $M$ and depends on the initial configuration $n$.

To study the thermal properties of the expansion dynamics, the thermal average over the ensemble of background spin states $\{\ket{\Psi^n}\}_{n=1...N}$ -- i.e. the ensemble of the Ising model at a given temperature $T$ -- must be performed. We achieve this using a standard Metropolis Monte Carlo algorithm to obtain a large number ($N=100$) of representative samples for desired temperatures $T$. For each of these samples $\ket{\Psi^n}$, the corresponding truncated base $\{\ket{\psi^n_\nu}\}_\nu$ is generated and the Schrödinger equation is solved on the restricted subspace starting from the initial state $\ket{\psi^n(\tau=0)}=\ket{\psi_1^n} \equiv \c_{\vec 0,\sigma_{\vec{0}}} \ket{\Psi^n}$.

Estimators for expectation values of observables such as the root mean square (rms) distance of the hole to the origin $r_\text{rms}$ can then be obtained by averaging the results obtained for each sample $n$,
\begin{equation}
r_\text{rms}(\tau)\approx  \frac{1}{N}\sum_{n=1}^N  \l\sum_{\vec{j}} \vec{j}^2 \bra{\psi^n(\tau)} \hat{n}^h_{\vec{j}} \ket{\psi^n(\tau)}\r^{1/2} 
\end{equation}
with $\hat{n}^h_{\vec{j}} = \prod_\sigma (1-\cd_{\vec{j},\sigma} \c_{\vec{j},\sigma})$ the hole density on site $\vec{j}$ which we evaluate in the truncated basis.

\emph{Numerical results.--}
In Fig.~\ref{fig1}b) we show typical numerically obtained time-traces of the hole's rms distance, for $t/J_z = 3$ and $M=10$. We observe slow spreading of the hole at low temperatures well below the Ising transition at $T_{\rm N}$, and faster spreading at high temperatures above $T_{\rm N}$. At longer times corresponding to a few tunneling events (typically we go up to times $\tau_{\rm max} = 15 / t$), both curves saturate. However, this is partly due to the finite dimension of the restricted basis we employ. 

We analyze the dependence of the long-time limit $r^{-1}_{\rm rms}(\tau_{\rm max})$ on the number of iterations $M$, corresponding to the maximum number of allowed tunneling events, in Fig.~\ref{fig1}c). For high temperatures, we observe scaling consistent with $r^{-1}_{\rm rms}(\tau_{\rm max}) \simeq M^{-1/2}$, i.e. the rms distance grows quickly and indefinitely. A scaling $\simeq M^{-1/2}$ with the square root of the number of allowed steps is expected from a classical random walk; this is true even at zero temperature for $J_z=0$, see Ref.~\cite{kanasznagy2017}. 

On the other hand, for low temperatures compared to $J_z$, we find $r^{-1}_{\rm rms}(\tau_{\rm max}) \to {\rm const.} > 0$ as $M^{-1/2} \to 0$, indicating slow spreading of the hole, bounded by a finite length scale $r_{\rm rms}^{\rm max}$ at time $\tau_{\rm max}$. We notice that for much longer times, on the order of $\tau_{\rm T}\gtrsim 100 /t$ \cite{Grusdt2018tJz}, Trugman loop effects are expected to lead to very slow but unbounded growth of $r_{\rm rms}$ \cite{Trugman1988}; however, these physics play no role on the time-scales up to $\tau_{\rm max}$ considered here.

Finally, we repeat the procedure described above for more values of the temperature $T$, in particular around the N\'eel transition temperature $T_{\rm N}=0.567 J_z$. The resulting extrapolated $r_{\rm rms}^{-1}(T;\tau_{\rm max}, M^{-1/2} \to 0)$ are plotted over temperature in Fig.~\ref{fig1}d). At a critical temperature around $T^* \approx 0.65 J_z$ close to but distinctly above $T_{\rm N}$, we find an abrupt change of behavior, with unbounded (bounded) growth of $r_{\rm rms}$ above (below) $T^*$. This is a main result of this Letter and, as discussed below, we interpret it as a dynamical signature of a confinement ($T<T^*$) to deconfinement ($T>T^*$) transition of the spin and charge sectors.

\begin{figure}[t!]
\centering
  \includegraphics[width=0.95\linewidth]{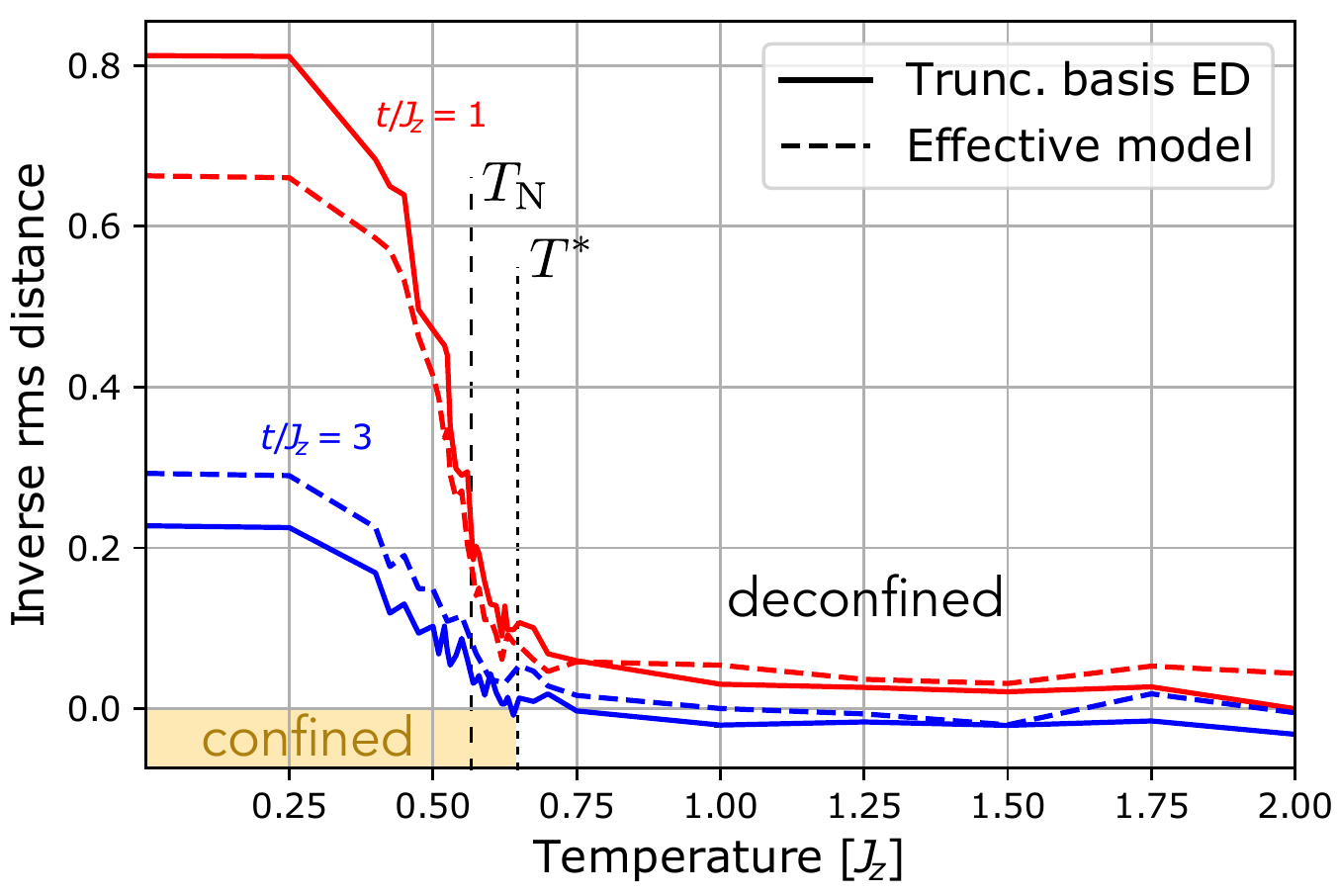}
\caption{We show long-time rms distances of a single hole extrapolated to $M^{-1/2} \to 0$ as a function of temperature, $r_{\rm rms}(T;\tau_{\rm max}, M^{-1/2} \to 0)$, for two ratios of $t/J_z=1$ (upper solid curve) and $t/J_z=3$ (lower solid curve). We compare our results to predictions by an effective spinon-chargon model (dashed curves) capturing the qualitative behavior. We indicate $T_{\rm N}$ and our estimate for $T^* = 0.65 J_z$ by vertical lines.}
\label{fig2}
\end{figure}

We performed a similar analysis as in Fig.~\ref{fig1} for a different value of $t/J_z=1$. The extrapolated long-time inverse rms distances are compared to the previous case in Fig.~\ref{fig2}. We find similar qualitative behavior, and remarkably the transition temperature $T^*$ does not change for different $t/J_z$. Overall the charge dynamics is only weakly affected by the spin background for $T>T^*$, while it depends strongly on the value of $J_z/t$ when $T<T^*$. These observations indicate a strikingly different interplay of spin and charge in the two regimes. 

At the given accuracy of our finite-size extrapolation $M^{-1/2} \to 0$, stating error bars on $T^*$ is challenging. We find our numerics most consistent with $T^* = 0.65(5) J_z$.

\emph{Effective parton model.--}
To obtain physical insight into our numerical results, we compare them with predictions by an effective parton model \cite{Coleman1984,Beran1996} of the $t-J_z$ model \cite{Grusdt2018tJz}. First we note that the initial creation of the hole changes both the spin- and charge quantum numbers, associated with the two global $U(1)$ symmetries of the $t-J_z$ model, by one; the initial state thus corresponds to a local spinon-chargon pair. 

In the subsequent dynamics, the chargon can move by distorting the surrounding spins. Since the Ising interaction is classical and generates no dynamics of its own, the spinon remains localized at the origin. Hence, the resulting spin configuration is determined entirely by the chargon's path; different paths may be assumed to be distinguishable up to self-retracing components, since they will lead to different spin configurations in the majority of the cases. I.e., the chargon motion effectively creates a memory of the hole's path through the spin background, in the form a of a (sometimes called geometric) string $\Sigma$ of displaced spins connecting the spinon to the chargon. At low temperatures, most strings lead to an increase of the net classical Ising energy $\H_{J_z}$, which acts as a potential energy, or string tension, for the chargon. 

Formally, in our effective parton model we replace the original $t-J_z$ Hilbertspace by a space spanned by orthogonal string states $\ket{\Sigma}$ with the spinon in the origin. States $\ket{\Sigma}$ in the effective Hilbert space correspond to unique states $\ket{\psi_\Sigma}$ in the $t-J_z$ Hilbertspace, but the opposite is not true. The effective Hamiltonian $\H_{\rm eff}$ consists of a tunneling term with amplitude $t$ between adjacent strings, and a potential energy term including the Ising interactions; see \cite{supp} for a detailed definition. At $T=0$, the dynamics obtained within this parton model is closely related to Brinkman and Rice's retraceable path approximation \cite{Brinkman1970}; at $T>0$ we average over thermal initial states $\ket{\Psi^n}$ as before.

Our intuitive physical picture above has its limitations. First, effects of loops are ignored; e.g., Trugman loops \cite{Trugman1988} and their generalizations to Ising configurations other than the N\'eel state effectively introduce spinon motion. Such processes are very slow and can be treated in a tight-binding approximation \cite{Grusdt2018tJz}. Second, not all physical states $\ket{\psi_\Sigma}$ are orthogonal for different $\Sigma$; in particular, if a plaquette along the path of the chargon has ferromagnetically aligned spins, paths along opposite directions around this plaquette are indistinguishable \cite{Carlstrom2016PRL,kanasznagy2017,Ji2020} and the corresponding quantum states have nonzero overlap. Our full numerical simulations introduced earlier systematically include these imperfections by constructing an orthonormalized restricted basis set. The number of iterations $M$ corresponds to the maximum string length $\ell_{\rm max}$ considered in the parton theory. 

In Fig.~\ref{fig2} we compare our earlier results to predictions by the effective parton model (dashed lines). We find qualitatively similar behavior, in particular the transition temperature $T^*$ is correctly captured by the effective model. This allows us to analyze the two qualitatively distinct dynamical regimes below and above $T^*$ within the simpler parton theory next.

\emph{Thermal spin-charge deconfinement.--}
Well below the Néel temperature, $T \ll T_N$, some spins will be thermally excited but magnetic order remains. Here the hole's movement is restrained in a similar way as for $T=0$ \cite{Bulaevskii1968}, resulting in confinement of the spinon and the chargon. Around $T_N$, the short-range correlations between the spins decrease rapidly, which may lead to a profound change in the behavior of the hole since these correlations provide a measure of the energy increase resulting from the chargon's movement. Specifically, the average energy $\langle \H_{J_z} \rangle_\ell$ of a string with length $\ell$, i.e. the string tension, is determined by local spin-correlations \cite{Grusdt2018tJz}. Notably, this does not imply that any change of behavior happens at exactly $T_N$, which is only a measure of long-distance correlations that do not directly affect the chargon's motion. Instead, the dynamical behavior changes at $T^* \neq T_{\rm N}$ in general (although $T^*$ and $T_{\rm N}$ are closely related).

We can estimate $T^*$ by considering the interplay of energy $E$ and entropy $S$ of string states in the effective parton model and ignoring quantum fluctuations $\propto t$. Taylor-expanding the string energy after averaging over the thermal spin-background and different string configurations with the same length $\ell$ allows us to write $E(\ell) \simeq \langle \H_{J_z} \rangle_\ell = E_0 + \ell \sigma + \ell^2 \sigma'/2  +...$. Assuming a microcanonical ensemble of strings, where all states of a given length $\ell$ are occupied equally, the entropy becomes $S \simeq \ell \log(z-1)$ where $z$ is the coordination number of the lattice ($z=4$ in the 2D square lattice we consider). Hence the free energy
\begin{equation}
  F=E-T S \simeq E_0 + \ell (\sigma - T \log(z-1))    
\end{equation}
is minimized for $\ell = 0$ (confined partons) when $T<T^*$ and for $\ell \to \infty$ (deconfined partons) when $T>T^*$. The thermal deconfinement transition takes place at
\begin{equation}
    T^* = \sigma / \log(z-1).
    \label{eqTscEst}
\end{equation}

As emphasized above, the string tension $\sigma$ depends only on the local spin correlations. Since these depend on temperature, Eq.~\eqref{eqTscEst} needs to be solved self-consistency for $T^*$ with $\sigma=\sigma(T^*)$. Using this procedure we predict $T^* = 0.65 J_z$, remarkably close to the observed value.

Another consequence of Eq.~\eqref{eqTscEst} is that $T^*$ becomes small in higher dimensions. In a $d$-dimensional hyper-cubic lattice $z=2 d$; since $\sigma = \mathcal{O}(T_{\rm N})$ is on the order of the N\'eel temperature, $T^* / T_{\rm N} \simeq 1/\log(d) \to 0$. Hence we expect that $T^*$ is systematically below $T_{\rm N}$ in high dimensions, further supporting our claim that the observed change of dynamical behavior at $T^*$ is not a mere reflection of the Ising transition at $T_{\rm N}$.

\emph{Thermalization dynamics.--}
Finally, we study how the mobile hole reaches a steady state when it spreads and interacts with the spin background. One would generically expect the isolated charge to equilibrate to a thermal state at the same temperature $T$ as the Ising spins. However, we observe pronounced deviations from this expected behavior.

\begin{figure}[t!]
\centering
  \includegraphics[width=0.85\linewidth]{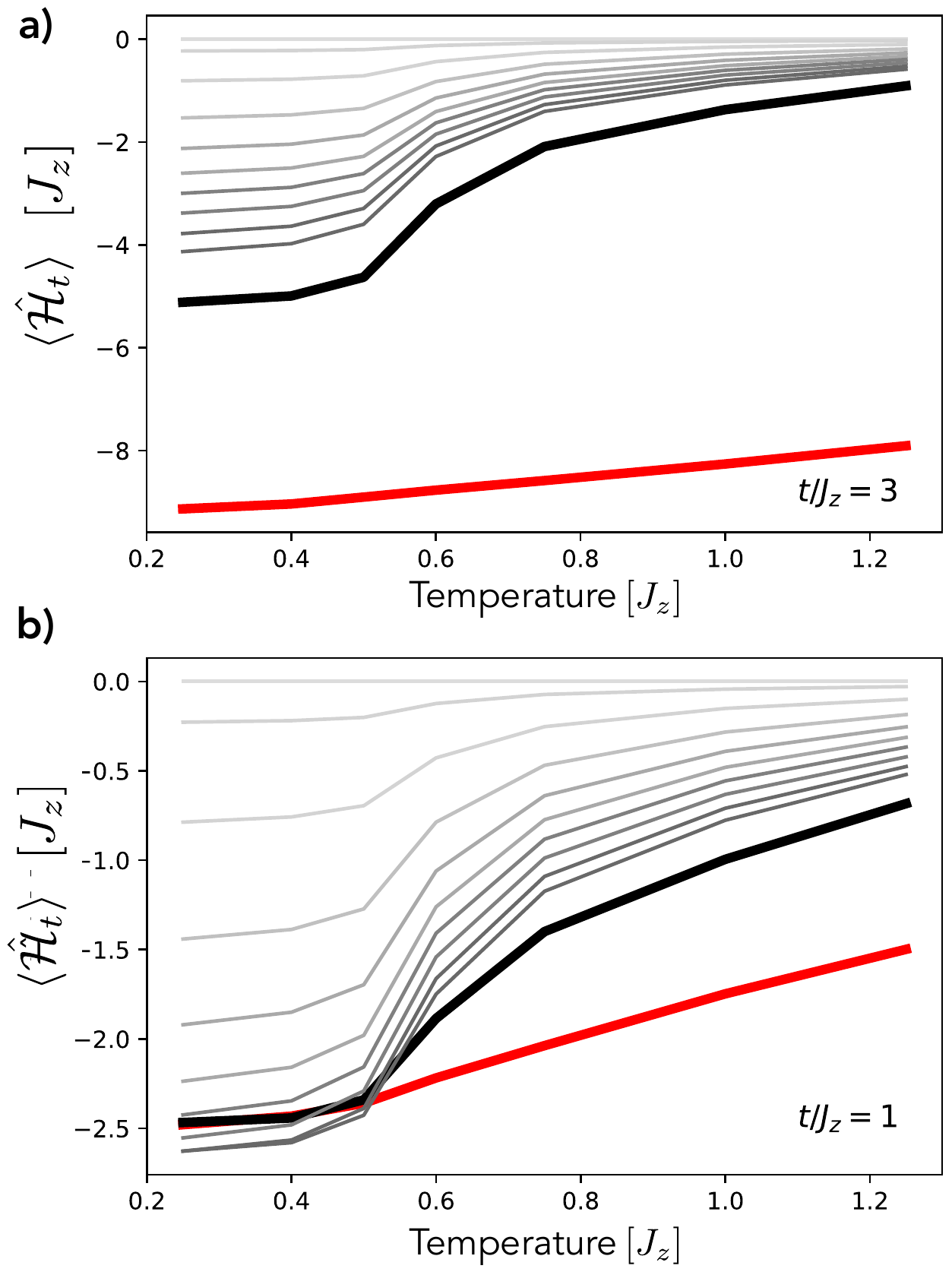}
\caption{The mobile dopant reaches a steady state (black) over a few tunneling times, as can be seen from its average kinetic energy $\langle \H_t \rangle$. We consider a) $t/J_z=3$ and b) $t/J_z=1$, and use the truncated-basis method described in the main text. The gray curves are the expectation value at earlier times, starting at $\tau=0$ and increasing in steps of $\Delta\tau=1/t$. Comparison of our results to a thermal ensemble (red) at the same temperature $T$ as the spin-background shows that the steady state is pronouncedly non-thermal in many cases.}
\label{fig3}
\end{figure}

In Fig.~\ref{fig3} we calculate the average kinetic energy $\langle \H_t \rangle$ of the hole, defining a local observable, which quickly relaxes to a steady state in a few tunneling times. Next we compare the steady-state result to a thermal ensemble at temperature $T$. To this end, we sample $n=1...N$ thermal background spin configurations $\ket{\Psi^n}$ as described above, introduce a hole, and apply a finite-temperature Lanczos method \cite{Jaklic1993} to describe the hole separately for each $n$. The thermal average of $N^{-1} \sum_n \langle \H_t \rangle_n$ over all samples $n$ is shown in Fig.~\ref{fig3}. 

At high temperatures, $T \gtrsim T^*$, the thermal ensemble deviates significantly from the steady state for both considered values of $t/J_z$. Within the effective parton model we attribute this behavior to the fact that the free energy is strongly dominated by the entropic contributions from a large number of long string states when $T>T^*$. Hence, in the post-quench dynamics the chargon can quickly populate these long-string states, which leads to the observed steady-state behavior. We expect much longer times would be required for the local kinetic energy to equilibrate too. We checked this picture by calculating string-length distributions and find that they quickly resemble the thermal ensemble \cite{supp}.

At low temperatures, $T \lesssim T^*$, we see in Fig.~\ref{fig3} that $\langle \H_t \rangle$ becomes thermal for $t/J_z=1$, whereas it remains non-thermal for larger $t/J_z=3$ at the considered times. In the latter case we believe that the significant separation of time scales in combination with the discrete spectrum of the Ising background leads to excessive thermalization times. To exchange energy with the spin environment, the chargon has to perform loops, which requires overcoming high energy barriers \cite{Trugman1988}. Indeed at low temperatures we find signatures for unoccupied loop states in the hole dynamics, which would be occupied in the thermal ensemble \cite{supp}.

\emph{Summary and Outlook.--}
We have established two temperature regimes $T \lessgtr T^*$ with distinct dynamical behavior of an initially localized hole moving in an Ising AFM. The observed dynamical transition at $T^*$ can be interpreted as thermal spinon-chargon deconfinement. While we cannot distinguish the deconfinement temperature $T^*|_{\rm 2D} \gtrsim T_{\rm N}$ from the N\'eel temperature $T_{\rm N}$ in our 2D simulations with absolute certainty, we expect from analytical arguments that $T^* < T_{\rm N}$ in higher dimensions. Further, we studied thermalization dynamics of a single hole in the $t-J_z$ model and revealed stable steady-states with non-thermal properties both in the confined and deconfined regimes.

Our theoretical analysis can be tested and extended experimentally using ultracold atoms in optical lattices \cite{Bohrdt2021PWA}. To realize the required Ising interactions, Rydberg dressing appears to be the most promising candidate. In particular this allows to realize AFM couplings for bosons \cite{Zeiher2016,Zeiher2017} or fermions \cite{GuardadoSanchez2020Ry}. For a single dopant the quantum statistics plays no role, extending the number of existing experimental setups that can address the quench dynamics studied in this Letter. Hence another possibility is to use spin-dependent interactions \cite{Duan2003,Trotzky2008,Dimitrova2019} to realize a bosonic model with AFM couplings. 

In the future, similar studies of the $SU(2)$ invariant $t-J$ model at finite temperature will be interesting. Experimentally, it is also conceivable to address hole dynamics at nonzero hole densities. Another interesting direction would be to explore thermalization dynamics of a single hole in the $t-J_z$ model at much longer times than addressed here. This may be possible using a combination of classical Monte Carlo sampling of the Ising background, as performed here, with large-scale time-dependent numerical DMRG (or tensor-network) simulations on extended cylinders \cite{Bohrdt2020_NJP,Hubig2020}.

\emph{Acknowledgements.--}
We thank E. Demler, U. Schollw\"ock, F. Palm, M. Knap, M. Greiner, M. Lebrat, G. Ji, M. Kanasz-Nagy, I. Lovas, Y. Wang, S. Ding, L. Rammelm\"uller, L. Pollet, I. Bloch and I. Cirac for fruitful discussions. The authors acknowledge funding from the European Research Council (ERC) under the European Union’s Horizon 2020 research and innovation programm (Grant Agreement no 948141), by the Deutsche Forschungsgemeinschaft (DFG, German Research Foundation) under Germany's Excellence Strategy -- EXC-2111 -- 390814868, by the NSF through a grant for the Institute for Theoretical Atomic, Molecular, and Optical Physics at Harvard University, and by the Smithsonian Astrophysical Observatory.


\newpage

\appendix

\section{Additional numerical results}

In this supplement, further numerical results obtained using the truncated-basis method are presented.

\begin{figure}[b!]
\centering
  \includegraphics[width=0.95\linewidth]{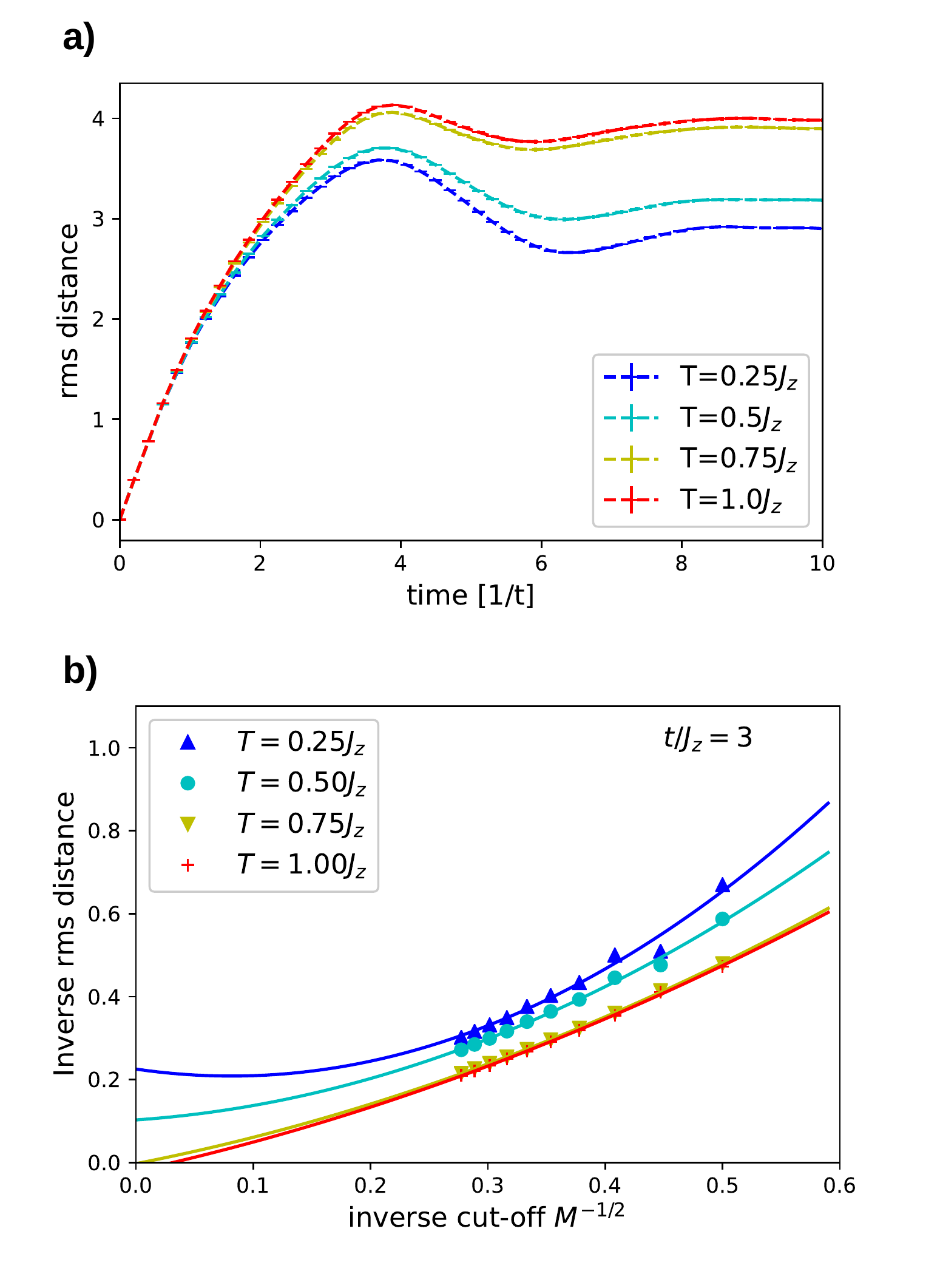}
\caption{This figure shows the data underlying Fig.~\ref{fig1} of the main text in more detail, i.e. the initial expansion of the hole, a), as well as its extrapolation towards an infinite system size ($M\to\infty$), b), both at $t/J_z=3$. In particular, one observes quantitatively similar behavior for high $T=0.75J_z$, $1.0J_z$, consistent with the prediction of a deconfined phase.}
\label{fig-a1}
\end{figure}

\subsection{Dynamical transition signatures}
In Fig.~\ref{fig-a1}, additional curves from the data underlying Fig.~\ref{fig1}d) are shown. Distinct behavior for low $T<T^*$ and high $T>T^*$ can easily be recognized after the initial ballistic expansion of the hole during the first few tunneling times. In particular, the hole shows similar behavior for all considered temperatures $T>T^*$. Indeed, in a deconfined phase the dynamics of the hole should be independent of $T>T^*$. While the rms distance might be expected to increase further over time in a deconfined phase, a saturation in the data can be explained by the finite size of the studied system. 

\begin{figure}[t!]
\centering
  \includegraphics[width=0.95\linewidth]{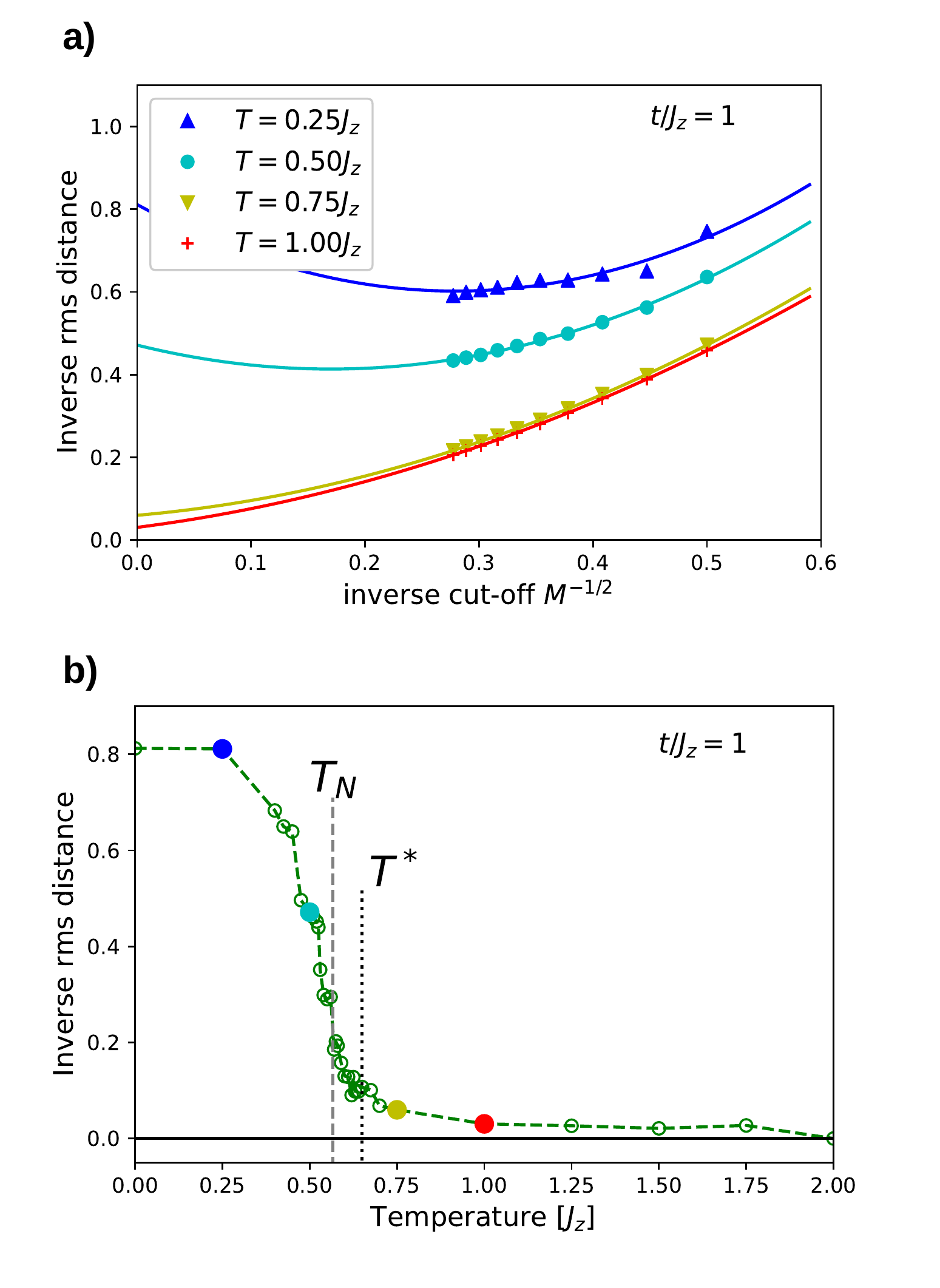}
\caption{Repeating the calculations of Fig.~\ref{fig1} in the main text for a lower value of $t/J_z=1$ reveals the transition temperature $T^*$ to be essentially independent of the value of $t/J_z$, as predicted by Eq.~\eqref{eqTscEst}. Here, the extrapolation to the thermodynamic limit $M\to\infty$, a), as well as the resulting inverse rms distances of the hole to the origin in the thermodynamic limit, b), are shown. In b) we indicate $T_{\rm N}$ and our estimate for $T^* = 0.65 J_z$ by dashed / dotted vertical lines.}
\label{fig-a2}
\end{figure}

These calculations can also be repeated for other values of $t/J_z$, see Fig.~\ref{fig-a2}, where the transition occurs around the same $T^*$ as before (within the capabilities of our finite-size extrapolation). Indeed, this indicates that the linear approximation of the string potential used in Eq.~\eqref{eqTscEst} successfully captures the essential phenomenology of the transition.

\begin{figure}[t!]
\centering
  \includegraphics[width=0.95\linewidth]{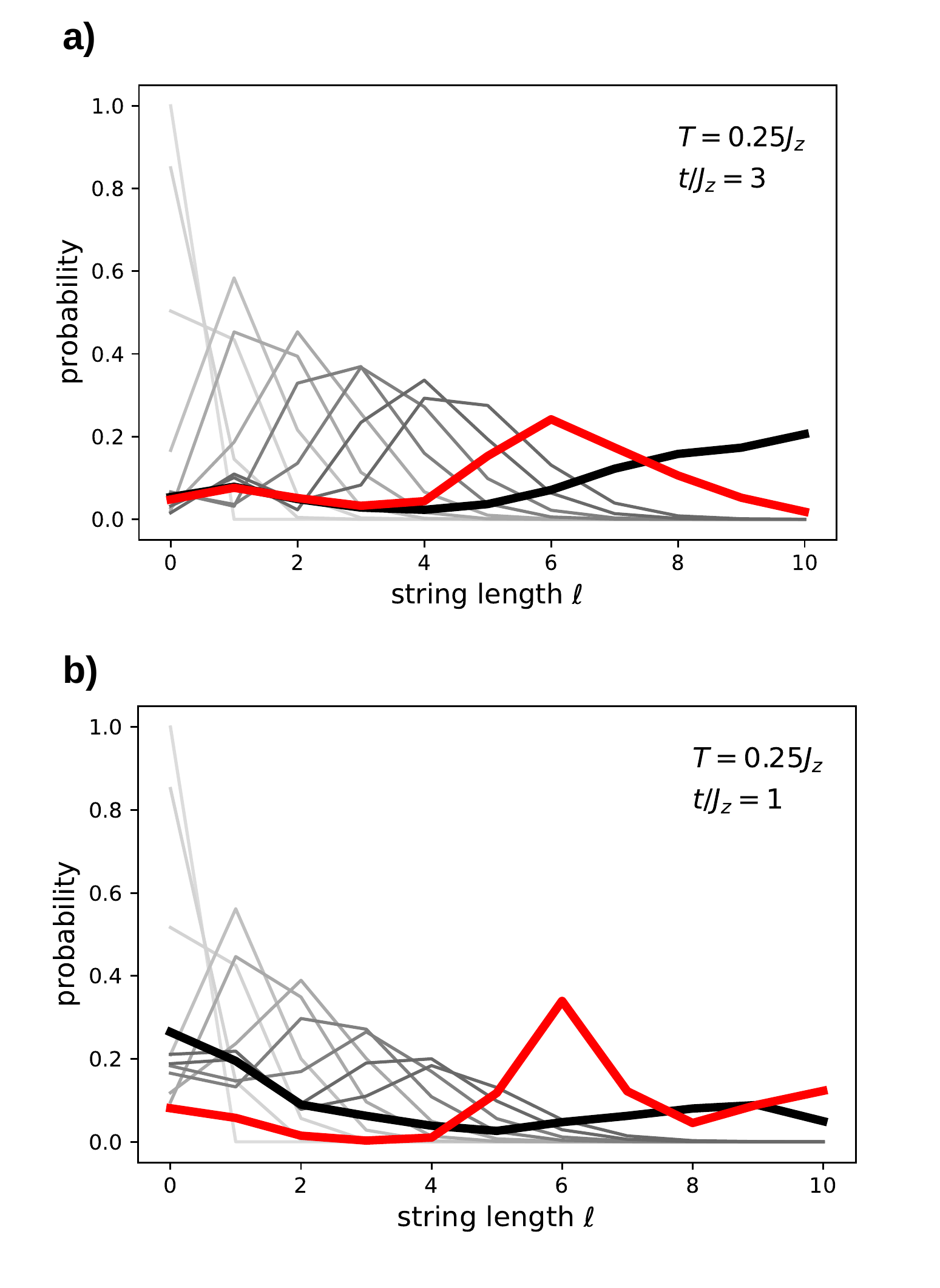}
\caption{The iteration step during which a state was added to the truncated basis may be associated with the string length of the corresponding state in the parton model. Insight into the thermalization dynamics may thus be gained by comparing the thermal string length distribution (red) with the steady state distribution (black) reached after several tunneling times, calculated using the truncated-basis method explained in the main text. Gray shaded curves correspond to distributions at earlier times $\tau$, starting at $\tau=0$ (lightest gray) and increasing in steps of $\Delta \tau = 1/t$.
Both the cases $t/J_z=3$, a), and $t/J_z=1$, b), show significant differences, here for $T=0.25J_z$. In particular, one observes a peak in the thermal distribution at $\ell=6$. This corresponds to the string length of the shortest possible Trugman loops. These states are not occupied in the steady state, despite their low energy. This indicates that much longer times are needed for the system to thermalize.}
\label{fig-a3}
\end{figure}

\subsection{Thermalization dynamics}
Similar to the comparison of the average kinetic energy in Fig.~\ref{fig3} of the main text, other observables may be used to investigate the thermalization dynamics of the mobile dopant in the $t-J_z$ model.

In particular, one may note that during the construction of the truncated basis, a state is added to the basis during the iteration step corresponding to the number of hops the hole has to perform to reach this state. Thus, the iteration step during which a state is added essentially corresponds to the string length in the spinon-chargon picture. By calculating the occupation probability for all states added during a given iteration $\ell = 0...M$, the distribution of string lengths $\ell$ may be reconstructed. The results can be seen in Fig.~\ref{fig-a3}.

It is noteworthy that the thermal distributions show clearly visible peaks at $\ell=6$. These peaks may be explained by the presence of Trugman loops \cite{Trugman1988} with low energy, which allow the hole to exchange energy with the spin lattice. The absence of these peaks in the long-time averages (obtained by averaging over the interval $[20/t,30/t]$) indicates that this exchange has not yet taken place and thus that much longer times are needed for the system to thermalize. 

\begin{figure*}[t!]
\centering
  \includegraphics[width=0.9\linewidth]{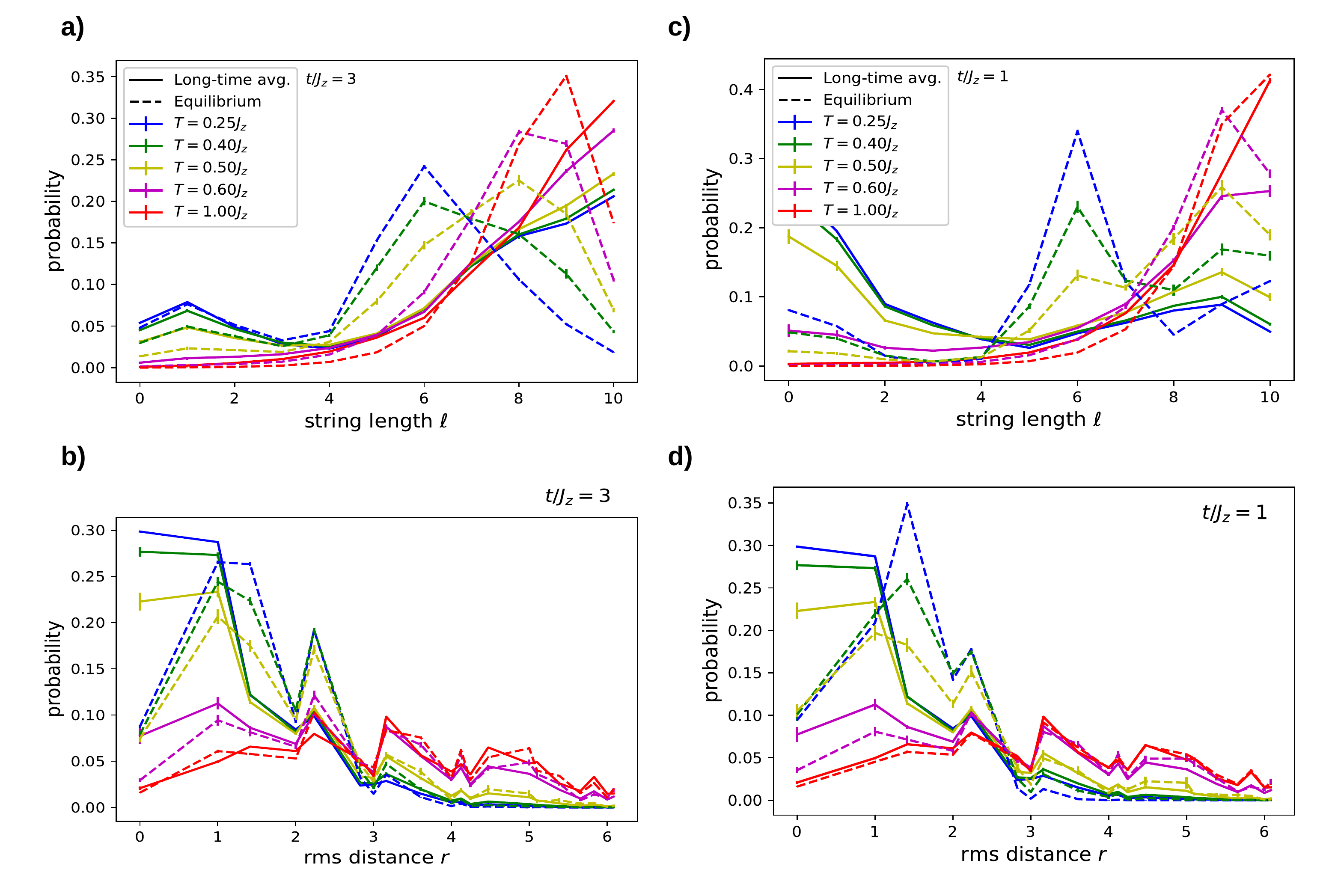}
\caption{Due to the numerically intractable size of the density matrix of the system at hand, the thermalization dynamics of the mobile dopant must be studied using suitable observables. Here, the string length distribution, a) as well as the hole's rms distance from the origin, b), are shown for $t/J_z=3$ [for $t/J_z=1$, c) and d), respectively]; legends in a) [in c)] also apply for b) [for d)]. Significant differences, in particular at low $T$, indicate that the system does not thermalize at the time scales considered in these calculations, going up to about 30 tunneling times.}
\label{fig-a4}
\end{figure*}

Similar conclusions can be taken from Fig.~\ref{fig-a4}, showing the string length distributions and rms of the hole density distributions for a wide range of temperatures across $T^*$. Interestingly, at high temperatures $T>T^*, T_{\rm N}$ these quantities are relatively close to their thermal values, in particular for the rms distance. We believe this can be explained by the fact that the rms distance and the string length-distributions in the deconfined regime reflect the delocalized hole and do not depend sensitively on its kinetic energy. The latter has not thermalized, as we demonstrated in Fig.~\ref{fig3} of the main text.

\section{Effective parton model calculations}
Here we provide details about our effective parton model calculations. These can be viewed as a simplification of the full truncated basis approach discussed in detail in the main text. As in the full truncated basis approach, we construct parton states labeled $\{ \ket{\Sigma^n} \}_\Sigma$ for a given spin background $\ket{\Psi^n}$ and perform thermal averages over $n=1...N$ Ising configurations afterwards.

\subsection{Physical intuition}
When considering the iteration process by which the truncated bases $\{ \ket{\psi^n_\nu} \}_\nu$ are generated in the full truncated basis approach -- successively applying $\H_t$, each time letting the hole hop one site further -- one may easily recognize a striking similarity of these bases to those resulting from a parton-based description of doped holes in antiferromagnets \cite{Bulaevskii1968,Grusdt2018tJz, Bohrdt2020_NJP}. Such parton theories consider the doped hole to be composed of two partons connected by a string of displaced spins: a heavy spinon and a light chargon, carrying the polaron's spin and charge degrees of freedom, respectively \cite{Beran1996}.

\subsection{Parton model Hilbertspace}
Since the spinon and the spin background are taken to be frozen due to the classical nature of the underlying Ising model, the chargon's path completely determines the resulting spin configuration; however only paths defined up to self-retracing components -- so-called strings $\Sigma$ -- are relevant to determine the final spin configuration. For a given initial spin configuration $\ket{\Psi^n}$, we define the string-zero state $\ket{\psi_1^n} := \c_{\vec{0},\sigma_{\vec{0}}} \ket{\Psi^n}$. Subsequent applications of individual hole hoppings, defining a string $\Sigma$, allows to construct string states $\ket{\psi_\Sigma^n}$ defined in the $t-J_z$ Hilbertspace $\mathscr{H}_{t-J_z}$. 

Importantly, the string states $\{ \ket{\psi_\Sigma^n} \}_\Sigma$ do \emph{not} define an orthonormal basis set. Loops and thermally excited spins may lead to some strings resulting in identical spin configurations, causing an overcompleteness of the string base $\{ \ket{\psi_\Sigma^n} \}_\Sigma$. This is explicitly taken into account in the full truncated basis approach, where out of the same set of states the orthonormal subset $\{\ket{\psi^n_\nu} \}_\nu$ is constructed. Significantly, the number of overcomplete states is relatively small for all considered temperatures if we consider AFM couplings $J_z > 0$. In the next step, this allows us to replace the original $t-J_z$ Hilbertspace $\mathscr{H}_{t-J_z}$ by an effective string Hilbertspace.

In mathematical terms, the effective string Hilbert space is defined as a tensor product
\begin{equation}
\mathscr{H}_{\rm eff}=\mathscr{H}_{\Sigma}\otimes\mathscr{H}_{\text{sp}}\otimes\mathscr{H}_{J_z}.
\end{equation}
Here $\mathscr{H}_{\Sigma}$ is the Hilbert space of all geometric strings connecting the chargon to the spinon; we postulate that the set $\{ \ket{\Sigma^n} \}_\Sigma$ defines an orthonormal basis, $\bra{\Sigma_1^n} \Sigma_2^n \rangle = \delta_{\Sigma_1,\Sigma_2}$. $\mathscr{H}_{\text{sp}}$ the Hilbert space of possible spinon positions on the lattice and $\mathscr{H}_{J_z}$ denotes the Hilbert space of possible spin configurations on the lattice before the hole is introduced into the system, i.e. the Hilbert space of the 2D Ising model. At this level of approximation, only the chargon is mobile, so only the dynamics in $\mathscr{H}_{\Sigma}$ will be considered. Note that while each string state $\ket{\Sigma^n}$ has a unique corresponding state $\ket{\psi^n_\Sigma}$ in the original $t-J_z$ Hilbertspace, the opposite is not true: different string states in $\mathscr{H}_{\rm eff}$, $\ket{\Sigma_{1}^n} \perp \ket{\Sigma_{2}^n}$, may correspond to the same $\ket{\psi^n_{\Sigma_1}} = \ket{\psi^n_{\Sigma_2}}$ in $\mathscr{H}_{t-J_z}$.

\subsection{Parton model Hamiltonian}
We derive the effective parton Hamiltonian by matching its matrix elements in the effective string Hilbertspace with the corresponding matrix elements of the original $t-J_z$ Hamiltonian in the $t-J_z$ Hilbertspace; i.e.:
\begin{equation}
    \bra{\Sigma_2^n} \H_{\rm eff}^{(n)} \ket{\Sigma_1^n} = \bra{\psi^n_{\Sigma_1}} \H \ket{\psi^n_{\Sigma_2}}.
    \label{eqHeffDef}
\end{equation}
We start by the hopping term. One easily confirms that the NN tunneling term $\H_{t}$ translates into a NN hopping term of equal amplitude between adjacent string states $\langle\Sigma_2,\Sigma_1\rangle$. Representing strings $\Sigma$ as sites of a Bethe lattice (as done e.g. in Ref.~\cite{Grusdt2018tJz}), $\H_t$ thus maps to an effective single-particle hopping problem on the Bethe lattice described by the effective Hamiltonian
\begin{equation}
\label{eq-Heff}
\H^t_{\rm eff}=-t\sum_{\langle\Sigma_2,\Sigma_1\rangle}\l\ket{\Sigma_2}\bra{\Sigma_1}+\text{h.c.}\r.
\end{equation}
Note that $\H^t_{\rm eff}$ is \emph{independent} of the index $n$ labeling the original spin background $\ket{\Psi^n}$ for which we construct the string states. This is a consequence of postulating that string states in $\mathscr{H}_{\rm eff}$ are orthonormal, and represents a key simplification as compared to the full truncated basis method. 

Next we construct $\H^{J_z,(n)}_{\rm eff}$ which defines a potential energy for the string. To this end, we only consider diagonal contributions in Eq.~\eqref{eqHeffDef}, with $\Sigma_1 = \Sigma_2$. This yields a string potential
\begin{equation}
    V^{(n)}_{\Sigma}=\bra{\psi_\Sigma^n}\H_{J_z}\ket{\psi_\Sigma^n}    
\end{equation}
corresponding to the energy of the lattice after the chargon has moved along the string starting from the given initial configuration $\ket{\Psi^n}$. It should be emphasized that $V^{(n)}_{\Sigma}$ depends not only on the string $\Sigma$, but also on the spinon position (taken to be fixed at the origin $\vec{j}=\vec{0}$ here) and the initial spin configuration $\ket{\Psi^n}\in\mathscr{H}_{J_z}$. 

The resulting potential energy in the effective Hamiltonian reads
\begin{equation}
    \H^{J_z,(n)}_{\rm eff} = \sum_{\Sigma}V_{\Sigma}^{(n)} \ket{\Sigma^n}\bra{\Sigma^n}.
\end{equation}
Together, the effective parton Hamiltonian becomes
\begin{equation}
    \H_{\rm eff}^{(n)} = \H^t_{\rm eff} +  \H^{J_z,(n)}_{\rm eff},
\end{equation}
where only the second term depends on $n$ on the right hand side of the equation.

\subsection{Thermal averages}
Similar to the truncated basis approach, thermal properties of the effective parton model can be accessed by using a Monte Carlo algorithm to sample initial spin states $\ket{\Psi^n}$. For each of the $n=1...N$ samples, the string potential $V_\Sigma$ can be calculated up to a maximum depth of the Bethe lattice $\ell_{\text{max}}$ at which the Hilbert space must be truncated due to computational limits. The results, e.g. dynamics as considered in our paper, can be calculated separately for each $n$ and be averaged over all $N$ samples in the end.

In our numerical simulations, we typically used $N=100$ and considered maximum string lengths of up to $\ell_{\rm max}=11$.

\subsection{Effective string potential}
As a concrete application, we calculate the \emph{average} string potential. I.e. we start from a thermal ensemble of $N$ Ising states $\ket{\Psi}^n$ at temperature $T$ and sample $S_\ell$ strings $\Sigma_\ell$ of a given length $\ell$ but with random orientations. Averaging over these string configurations yields
\begin{equation}
    V(\ell,T) = \frac{1}{N} \sum_{n=1}^N \frac{1}{S_\ell} \sum_{\Sigma_\ell} V^{(n)}_{\Sigma_\ell}.
\end{equation}

\begin{figure}[t!]
\centering
  \includegraphics[width=0.85\linewidth]{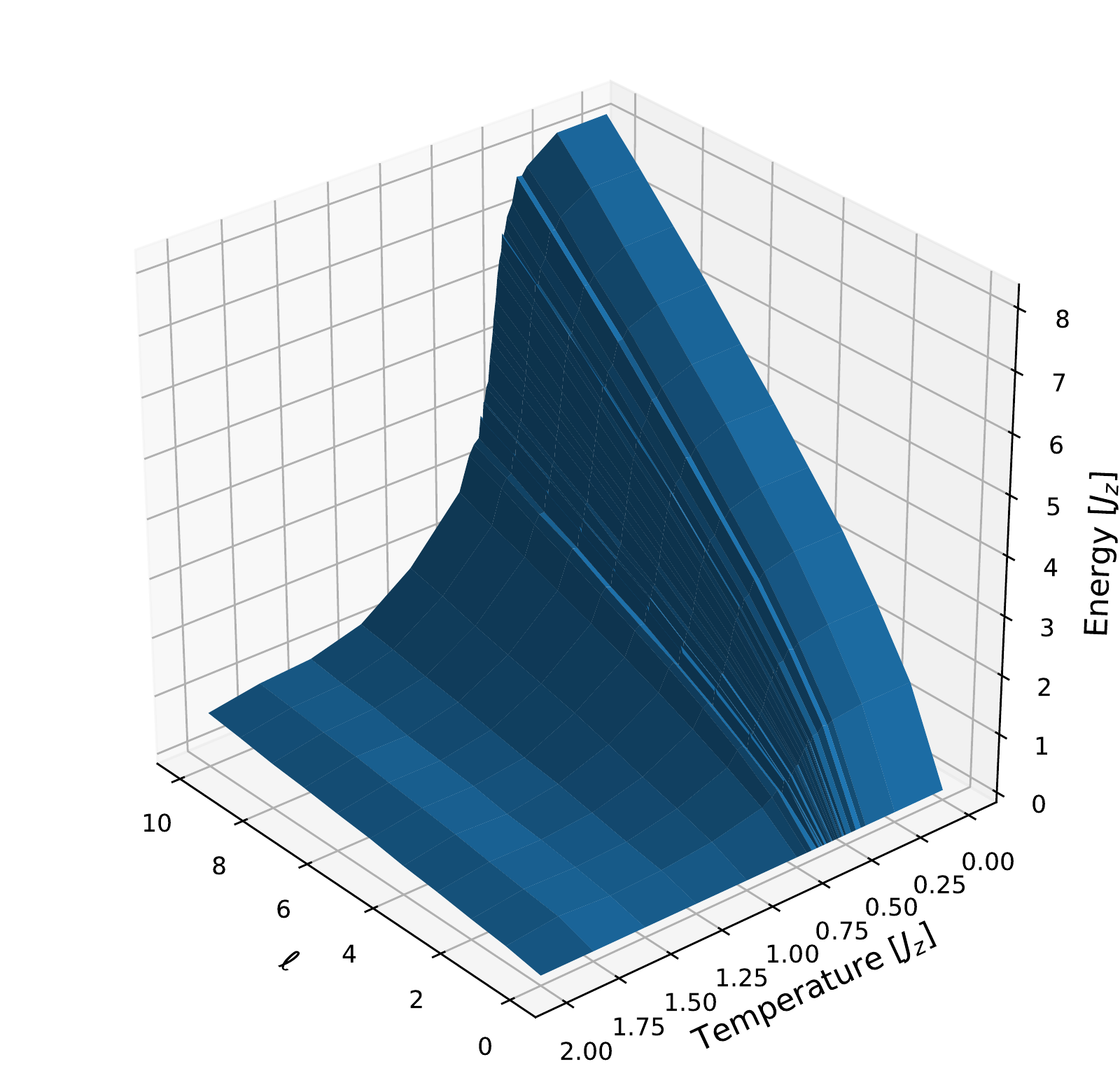}
\caption{By sampling both thermal states of the Ising system as well as strings of varying length, the potential acting on the strings can be calculated as a function of the temperature $T$. At fixed $T$ the potential is approximately linear everywhere.}
\label{fig-b1}
\end{figure}

The results are shown in Fig.~\ref{fig-b1}, showing the string potential to be approximately linear $\simeq \sigma \ell$ in $\ell$ both for temperatures below and above $T^*$. Around $T \approx 0.5 J_z$, close to $T_{\rm N}$ and $T^*$, a sharp drop of the linear string tension $\sigma$ can be observed.

\end{document}